\documentclass[12pt,preprint]{aastex} 
\catcode`\@=11
\def\gsim{\ifmmode{\mathrel{\mathpalette\@versim>}}
    \else{$\mathrel{\mathpalette\@versim>}$}\fi}
\def\lsim{\ifmmode{\mathrel{\mathpalette\@versim<}}
    \else{$\mathrel{\mathpalette\@versim<}$}\fi}
\def\@versim#1#2{\lower 2.9truept \vbox{\baselineskip 0pt \lineskip
    0.5truept \ialign{$\m@th#1\hfil##\hfil$\crcr#2\crcr\sim\crcr}}}
\catcode`\@=12

\def\mstar{M_\ast}
\def\mgas{M_g}
\def\mbh{M_{BH}}
\def\kinstar{K_\ast}

\def\intengas{U_g}
\def\potstar{\phi_\ast}
\def\potgas{\phi_g}
\def\rhostar{\rho_\ast}
\def\sigstar{\sigma_\ast}
\def\fstarmean{\langle f_\ast\rangle}
\def\rhogas{\rho_g}
\def\sigv{\sigma_v}
\def\sigvsq{\sigv^2}
\def\rv{r_v}
\def\Kv{K_v}

\def\mstarone{M_{\ast 1}}
\def\mgasone{M_{g1}}
\def\kinstarone{K_{\ast 1}}

\def\sigvonesq{\sigma_{v1}^2}
\def\rvone{r_{v1}}

\def\mstartwo{M_{\ast 2}}
\def\mgastwo{M_{g2}}
\def\kinstartwo{K_{\ast 2}}

\def\sigvtwosq{\sigma_{v2}^2}
\def\rvtwo{r_{v2}}

\def\mstarz{M_{\ast 0}}
\def\mgasz{M_{g0}}

\def\mstari{M_{\ast i}}
\def\mgasi{M_{gi}}

\def\re{R_{\rm e}}
\def\ML{\Upsilon_\ast}
\begin{document}
\title{The importance of dry and wet merging \\
on the formation and evolution of elliptical galaxies}

\author{L. Ciotti$^1$, B. Lanzoni$^2$, M. Volonteri$^3$}
\affil{$^1$ Dipartimento di Astronomia, Universit\`a di Bologna, 
            via Ranzani 1, 40127 Bologna, Italy \\
       $^2$INAF -- Osservatorio Astronomico di Bologna, via Ranzani 1,
            40127, Bologna, Italy \\ 
       $^3$ Institute of Astronomy, 
            University of Cambridge, Madingley Road, UK\\}

\date{November 10, 2006, accepted version}

\begin{abstract}
With the aid of a simple yet robust approach we investigate the
influence of dissipationless and dissipative merging on galaxy
structure, and the consequent effects on the scaling laws followed by
elliptical galaxies. Our results suggest that ellipticals cannot be
originated by parabolic merging of low mass spheroids only, even in
presence of substantial gas dissipation. However, we also found that
scaling laws such as the Faber-Jackson, Kormendy, Fundamental Plane,
and the $\mbh -\sigma$ relations, when considered over the whole mass
range spanned by ellipticals in the local universe, are robust against
merging.  We conclude that galaxy scaling laws, possibly established
at high redshift by the fast collapse in pre-existing dark matter
halos of gas rich and clumpy stellar distributions, are compatible
with a (small) number of galaxy mergers at lower redshift.
\end{abstract}

\keywords{galaxies: elliptical and lenticulars, cD -- 
          galaxies: evolution -- 
          galaxies: formation -- 
          galaxies: structure --
          galaxies: fundamental parameters}

\section{Introduction} 
\label{sec:intro}
 
Early-type galaxies are known to follow well defined empirical scaling
laws relating their global observational properties, such as total
luminosity $L$, effective radius $\re$, and central velocity
dispersion $\sigma$.  Among others we recall the Faber \& Jackson
(1976, hereafter FJ), Kormendy (1977), Fundamental Plane (Djorgovski
\& Davis 1987; Dressler et al. 1987; hereafter FP), the color-$\sigma$
(Bower, Lucey \& Ellis 1992), and the Mg$_2-\sigma$ (e.g., Guzman et
al. 1992; Bernardi et al. 2003c) relations.  In addition, it is now
believed that all elliptical galaxies host a central supermassive
black hole (SMBH; e.g., see de Zeeuw 2001), whose mass $\mbh$ scales
with the stellar mass $\mstar$ and velocity dispersion $\sigma$ of the
host galaxy (Magorrian et al. 1998; Ferrarese \& Merritt 2000; Gebhardt et
al. 2000; Tremaine et al. 2002).  Clearly, these scaling relations
provide invaluable information about the formation and evolution of
early-type galaxies, and set stringent constraints to galaxy formation
models.

The two major formation models for ellipticals that have been proposed
so far are the monolithic (Eggen, Lynden-Bell \& Sandage 1962), and
the merging (Toomre 1977; White \& Frenk 1991) scenarios. Each of them
scores observational and theoretical successes and drawbacks (e.g.,
see Ostriker 1980, Renzini 2006). For instance, we list here three
observational and theoretical evidences in favour of a fast and
dissipative monolithic-like collapse.

First, the observed color-magnitude and Mg$_2$-$\sigma$ relations, and
the increase of the $[\alpha/{\rm Fe}]$ ratio with $\sigma$ in the
stellar population of elliptical galaxies (e.g., see J{\o}rgensen
1999, Thomas, Greggio \& Bender 1999 Saglia et al. 2000; Bernardi et
al. 2003c, and references therein), suggest that star formation in
massive ellipticals was not only more efficient than in low mass
galaxies, but also that it was a faster process (i.e., completed
before SNIa explosions take place), with the time scales of gas
consumption and ejection shorter or comparable to the galaxy dynamical
time (e.g., see Matteucci 1994, Pipino \& Matteucci 2004), and
decreasing for increasing galaxy mass.

Second, structural and dynamical properties of ellipticals are well
reproduced by cold dissipationless collapse, a process which is
expected to dominate the last stages of highly dissipative collapses,
in which the gas cooling time of the forming galaxy is shorter than
its dynamical time, so that stars form `in flight', and the subsequent
dynamical evolution is a dissipationless collapse. It is now well
established that the end-products of cold and phase-space clumpy
collapses have projected density profiles well described by the
$R^{1/4}$ de Vaucouleurs (1948) law, radially decreasing line-of-sight
velocity dispersion profiles, and radially increasing velocity
anisotropy, in agreement with what observed in elliptical galaxies
(e.g., see van Albada 1982; May \& van Albada 1984; McGlynn 1984;
Aguilar \& Merritt 1990; Londrillo, Messina \& Stiavelli 1991; Udry
1993; Hozumi, Burkert, \& Fujiwara 2000; Trenti, Bertin \& van Albada
2005).  

Third, the current and remarkably succesful cosmological scenario for
structure formation predicts that well defined scaling laws are
imprinted in the dark matter halos; in particular, the virial velocity
dispersion of DM halos increases as $\sigv\propto M_{DM}^{1/3}$. This
because virialized DM halos are the \emph{collapse} end-products of
negative energy (inhomogeneous) density distributions, in which the
absolute value of the binding energy per unit mass increases with the
halo mass (Peebles 1980). On the contrary, in a parabolic
merging $\sigv$ would not increase with halo mass.

Thus the observed scaling laws of elliptical galaxies could be
originated by the fast collapse of inhomogeneous gas and star
distributions in pre-existing DM halos, rather than by parabolic
merging processes (e.g., see Lanzoni et al. 2004). Note that high
resolution N-body simulations (Nipoti, Londrillo \& Ciotti 2006) have
shown that cold (dissipationaless) collapses in pre-existing DM halos
nicely reproduce the weak homology of elliptical galaxies (e.g., see
Caon, Capaccioli \& D'Onofrio 1993; Prugniel \& Simien 1997; Bertin,
Ciotti \& Del Principe 2002, hereafter BCD02; Graham \& Guzm\'an 2003)
and the central break radius in their surface brightness profile
(Ferrarese et al. 1994; Lauer et al. 1995; Graham et al. 2003;
Trujillo et al. 2004).

The last point above is particularly puzzling because the available
observations seem to indicate that mergers may happen in the life of
elliptical galaxies, with dissipative (wet) mergers dominating at high
redshift, and gas-free (dry) merging mainly affecting massive
ellipticals at $z\lsim 1.5$ (e.g., see Bell et al. 2004, 2006; van
Dokkum 2005; Faber et al. 2005; Conselice 2006). This picture is
also suggested by the available information on the star formation
history of the Universe and the redshift evolution of the quasar
luminosity function (see, e.g., Haehnelt \& Kauffmann 2000; Burkert \&
Silk 2001; Yu \& Tremaine 2002; Cavaliere \& Vittorini 2002; Haiman,
Ciotti \& Ostriker 2004).  In addition, parabolic orbits seem to be
quite relevant in the hierarchical merging picture (e.g., see Benson
2005; Khochfar \& Burkert 2006).  To get insights on this issue, in
the present paper we will focus on the remarkable homogeneity and
regularity of the family of early-type galaxies (testified by their
scaling laws), and we explore the consequences of galaxy merging on
them.

The impact of dry merging on the scaling laws of early-type galaxies has
been already investigated in several works (e.g., Capelato, de
Carvalho \& Carlberg 1995; Pentericci, Ciotti \& Renzini 1996;
Haehnelt \& Kauffmann 2000; Ciotti \& van Albada 2001, hereafter
CvA01; Evstigneeva, Reshetnikov \& Sotnikova 2002; Nipoti, Londrillo
\& Ciotti 2003, hereafter NLC03; Gonz\'alez-Garc\'{\i}a \& van Albada
2003; Dantas et al. 2003; Evstigneeva et al. 2004;
Boylan-Kolchin, Ma \& Quataert 2005, 2006).  In particular, the simple
approach of CvA01 and the N-body simulations of NLC03 showed that
repeated, parabolic merging of gas-free galaxies is unable at
reproducing the observed scaling laws, because the merger-products are
characterized by unrealistically large effective radius and
mass-independent velocity dispersion (see also Shen et al. 2003).
However, simple physical arguments show that gas dissipation should be
able to mitigate the problems posed by dry merging to the explanation
of the observed scaling laws (e.g., see CvA01; Kazantzidis et
al. 2006; Robertson et al. 2006ab; Dekel \& Cox 2006)\footnote{Note that
by comparing the FP of galaxies and that of galaxy clusters, Burstein
et al. (1997) and Lanzoni et al. (2004) suggested that, at variance
with groups and clusters, gas dissipation must have had an important
role on the formation and evolution of ellipticals.}.  Unfortunately,
numerical simulations with gas dissipation are considerably more
complicate than pure N-body simulations (e.g., see S\'aiz,
Dom\i\'nguez-Tenreiro \& Serna 2004; O\~norbe et al. 2005, 2006; Robertson
et al. 2006ab), and, in particular, very few of them have been done from
realistic cosmological initial conditions (e.g., see Naab et al. 2006,
and references therein).  For these reasons, by generalizing the
approach presented in CvA01 to the dissipative case, we further
investigate with Monte-Carlo simulations the compatibility of galaxy
merging with the formation and evolution of early-type galaxies,
focusing in particular on 1) the effects of gas dissipation on the
merger end-products (wet merging), and 2) the effects of parabolic dry
and wet merging on the scaling laws followed by elliptical galaxies in
the local universe.  We argue that parabolic merging of low-mass seed
galaxies alone cannot be at the origin of the scaling laws, even
though wet mergers lead to early-type galaxies following the observed
scaling laws better than the end products of dry merging. We also show
that galaxy scaling laws, such as the FJ, Kormendy and FP relations,
once in place, are robust against merging. Thus, our results reinforce
the idea that monolithic-like collapse at early times and subsequent
merging could just represent the different phases of galaxy formation
(collapse) and evolution (merging, in addition to the aging of the
stellar populations and related phenomena).

The paper is organized as follows. In Section \ref{sec:models} we
derive the recursive equations describing the evolution of galaxy
properties after dry and wet parabolic mergers, and in Section 3 we
discuss in detail the case of equal-mass merging.  In Section
\ref{sec:simu} we use the derived relations in Monte-Carlo
investigations of merging of elliptical galaxy populations, and the
main results are finally summarized and discussed in Section
\ref{sec:concl}.

\section{The models}
\label{sec:models}
In this Section we now derive from elementary physics arguments the
relations between the properties of the progenitor galaxies and those
of the merging end-products that will be used in the rest of the
paper. For simplicity in the adopted scheme each elliptical is modeled
as a non rotating, isotropic and spherically symmetric virialized
system, characterized by a stellar mass $\mstar$, a gas mass $\mgas =
\alpha \mstar$, and a SMBH mass $\mbh = \mu \mstar$; from observations
$\mu\simeq 10^{-3}$ in $z=0$ spheroids (Magorrian 1998). In our
treatment we do not consider the presence of a DM halo, as it could be
introduced just by rescaling the model stellar mass-to-light ratio if
the DM density distribution is proportional to the stellar one, as
discussed in the following paragraphs.  The total energy of a galaxy
is then given by
\begin{equation}
E=\kinstar +\intengas +W,
\label{eq:Etot}
\end{equation}
where 
\begin{equation}
\kinstar = {3\over 2} \int{\rhostar\,\sigstar^2\, dV}
\label{eq:kinstar}
\end{equation}
is the stellar kinetic energy, and 
\begin{equation}
\intengas = \frac{3\,k_B}{2\,\langle m\rangle} \int{\rhogas\,T\, dV}
\label{eq:intgas}
\end{equation}
is the gas internal energy; $\sigstar$, $k_B$, $T$, and $\langle
m\rangle$ are the stellar 1-dimensional velocity dispersion, the
Boltzmann constant, the gas temperature, and the gas mean molecular
mass, respectively. Finally
\begin{equation}
W = {1\over 2} \int{(\rhostar+\rhogas) (\potstar+\potgas) \,dV}
\label{eq:W}
\end{equation}
is the total gravitational energy of stars and gas (we do not consider the
negligible contribution of the central SMBH).

Under the simplifying assumption that the gas is spatially distributed
as the stars (i.e., $\rhogas=\alpha\rhostar$), then $\potgas=\alpha\,
\potstar$, and
\begin{equation}
W = (1+\alpha)^2 \, W_\ast, 
\label{eq:W2}
\end{equation}
where $W_\ast$ is the self-gravitational energy of the stellar
component.  Furthermore, assuming that the gas is in equilibrium in
the total gravitational field, from the Jeans and the hydrostatic
equations it results that $T=\langle m\rangle \sigstar^2/k_B$, and
from equations (\ref{eq:kinstar})-(\ref{eq:intgas})
\begin{equation}
\intengas = \alpha\kinstar.
\label{eq:Ug}
\end{equation}
Finally, from equations (\ref{eq:W2})-(\ref{eq:Ug}) and the virial
theorem for the two-component system of stars + gas, the total
galactic energy can be written in terms of the stellar energy and of
the relative amount of gas as
\begin{equation} 
E = -(1+\alpha)\kinstar ={(1+\alpha)^2\over 2} W_\ast .  
\label{eq:EtVT} 
\end{equation}
Note that a DM halo of mass $M_{\rm DM}=\alpha_{\rm DM}\mstar$
distributed as the stars would be easily considered in the present
scheme by the addition of the new parameter $\alpha_{DM}$ in equations
(\ref{eq:W2})-(\ref{eq:EtVT}).

The quantities introduced so far are not directly observables, and so
in Section 2.1 we will show how to relate the characteristic
one-dimensional stellar velocity dispersion $\sigv$ and the
characteristic radius $\rv$, defined as
\begin{equation}
\kinstar \equiv {3\over 2} \,\mstar \,\sigvsq,
\label{eq:Tstar}
\end{equation}
\begin{equation}
|W_\ast| \equiv {G\, \mstar^2\over\rv},
\label{eq:Wstar}
\end{equation}
to the galaxy effective radius $\re$ and central projected velocity
dispersion $\sigma$\footnote{Note that $\sigv$ and $\rv$ in equations
(\ref{eq:Tstar}) and (\ref{eq:Wstar}) coincide with the virial
velocity dispersion and the virial radius of the star$+$gas
system. This would not be true in a system where $\rho_*\neq\rho_g$.}.

We now focus on the parabolic merging of two galaxies, so that the
total energy of the system is the sum of the internal potential and
kinetic energies of the two progenitor galaxies; we also assume that
no mass is lost in the process. During the merging, as a consequence
of gas dissipation, a fraction $\eta$ of the available gas mass is
converted into stars, and the stellar mass balance equation is
\begin{equation}
\mstar = \mstarone +\mstartwo +\eta (\mgasone +\mgastwo).
\label{eq:mstar}
\end{equation}
Furthermore, a new SMBH forms by the coalescence of the two central
BHs and a fraction $f\eta$ of the available gas is accreted on it,
leading to a BH of final mass
\begin{equation}
\mbh = (M_{BH1}^p +M_{BH2}^p)^{1/p} +f\eta (\mgasone +\mgastwo).
\label{eq:mbh}
\end{equation}
The free parameter $1\le p\le 2$ describes how much BH rest mass is
radiated as gravitational waves during the BH coalescence: $p=1$
corresponds to the classical merging case (no gravitational
radiation), while $p=2$ to the maximally radiative case for
non-rotating BHs.  Note that in equation (\ref{eq:mbh}) it is
implicitly assumed that first $M_{BH1}$ and $M_{BH2}$ merge, and then
the gas is accreted on the new BH; the other extreme case would be
that of gas accretion followed by merging (e.g., see Hughes \&
Blandford 2003). Of course, if $p=1$ there is no difference in the
final mass; we anticipate that in the Monte-Carlo simulations
described in Section \ref{sec:simu} we explored both cases, finding not
significative differences.  As a consequence of star formation and BH
accretion, the gas mass balance equation is
\begin{equation}
\mgas = (1-\eta -f\eta)(\mgasone +\mgastwo), 
\label{eq:mgas}
\end{equation}
which implies that $0\le\eta\le 1/(1+f)$. Thus, the gas-to-star mass
ratio after the merger and the new Magorrian coefficient are given by
\begin{equation}
\alpha \equiv \frac{\mgas}{\mstar} = \frac{(1-\eta-f\eta) (\alpha_1\mstarone
+\alpha_2\mstartwo)}{(1+\eta\alpha_1)\mstarone +(1+\eta\alpha_2)\mstartwo},
\label{eq:alpha}
\end{equation}
while
\begin{equation}
\mu \equiv \frac{\mbh}{\mstar} = \frac{(\mu_1^p\,M_{\ast 1}^p
+\mu_2^p\,M_{\ast 2}^p)^{1/p} +f\eta\, (\alpha_1\mstarone+\alpha_2\mstartwo)}
{(1+\eta\alpha_1)\mstarone +(1+\eta\alpha_2)\mstartwo},
\label{eq:mu}
\end{equation}
respectively. Note that if $p=1$ and $f=\mu_1=\mu_2$, the
proportionality coefficient $\mu$ remains unchanged after the merging;
also note that the scheme above is generalizable by allowing for
different values of $f$ and $\eta$ in the two progenitor galaxies,
but for simplicity in this paper we assume $f$ and $\eta$ fixed.

In order to describe the effects on $\rv$ and $\sigv$ of the radiative
energy losses associated with gas dissipation, a fraction $(1+f)\eta$
of the gas internal energy $U_g$ of each progenitor is subtracted from
the total energy budget of the merger-product, consistently with the
previous assumptions\footnote{This represents the limit case where
energy losses affect the internal energies of the progenitor galaxies
\emph{before} they merge. In the other limit case the two galaxies
would merge without dissipation, and then a fraction $(1+f)\eta$ of
the resulting total gas mass and of the internal energy $U_0=-\alpha_0
(E_1 +E_2)/(1+\alpha_0)$ would be dissipated, where
$\alpha_0=(\mgasone +\mgastwo)/(\mstarone +\mstartwo)$. The two
schemes lead to identical predictions when $\alpha_1 =\alpha_2$, or,
for $\alpha_1\neq\alpha_2$, when $\sigma_1=\sigma_2$.}. 
Thus, from equation (\ref{eq:Ug}), the final total energy of the
remnant is
\begin{equation}
E = E_1 + E_2 -\eta\,(1+f)\,(\alpha_1\kinstarone +\alpha_2\kinstartwo).
\label{eq:Ep}
\end{equation}
Identity (\ref{eq:EtVT}) with the new total energy $E$, the new mass
ratio $\alpha$ and the new total stellar mass $\mstar$ are given by
equations (\ref{eq:Ep}), (\ref{eq:alpha}), and (\ref{eq:mstar}),
respectively. Simple algebra shows that for the new galaxy
\begin{equation}
\sigvsq = {\mstarone+\mgasone\over \mstar+\mgas}A_1\sigvonesq +
          {\mstartwo+\mgastwo\over \mstar+\mgas}A_2\sigvtwosq,
\label{eq:sigp}
\end{equation}
and
\begin{equation}
{1\over\rv} = \left({\mstarone+\mgasone\over \mstar+\mgas}\right)^2 
              {A_1\over\rvone}+ 
              \left({\mstartwo+\mgastwo\over \mstar+\mgas}\right)^2 
              {A_2\over\rvtwo},
\label{eq:rp}
\end{equation}
where 
\begin{equation}
A_1 = 1 + {(1+f)\eta\alpha_1\over 1+\alpha_1},
\label{eq:Ai}
\end{equation}
and a similar expression holds for $A_2$. In a dry ($\eta=0$) merging
$A_1=A_2=1$, so that
\begin{equation}
{\rm min} (\sigvonesq, \sigvtwosq)\le\, \sigvsq =
\frac{(1+\alpha_1)\mstarone\sigvonesq + (1+\alpha_2)\mstartwo\sigvtwosq}
{(1+\alpha_1)\mstarone + (1+\alpha_2)\mstartwo} \le {\rm max} (\sigvonesq,
\sigvtwosq),
\label{eq:sigv}
\end{equation}
i.e., the virial velocity dispersion of the merger-product cannot be
larger than the maximum velocity dispersion of the progenitors (the
$\alpha>0$ and $\eta=0$ case also describes the situation in which the
gaseous component is replaced by a DM halo). Instead, $A>1$ in case of
wet ($\eta>0$) merging, and the resulting $\sigv$ is larger than in
the dry case, possibly larger than the maximum velocity dispersion of
the progenitors.  A similar argument shows that in presence of gas
dissipation the new $\rv$ increases less than in the dry case. Note
that the conclusions of this preparatory analysis are obtained under
the hypothesis of parabolic merging. If mergers involve galaxies on
bound orbits, the additional negative energy term in equation
(\ref{eq:Ep}) would lead to an increase of $\sigv$ also in equal-mass
dry mergers.  The analysis of this case, and the question of how much
fine-tuned the properties of the progenitor galaxies should be with
their binding orbital energy in order to reproduce the scaling laws,
are not further discussed in this paper (e.g., see Boylan-Kolchin et
al. 2005, 2006; Almeida, Baugh \& Lacey 2006).

\subsection{Relating intrinsic to observational properties: weak homology 
effects}
So far the discussion involved galaxy virial properties only. However,
galaxy scaling laws relate observational quantities such as total
luminosity $L$, central projected velocity dispersion $\sigma$
(luminosity averaged over some aperture), and circularized effective
radius $\re$. For example, in this paper we compare our models with
the FJ ($L \propto \sigma^{3.92}$, rms[$\log\sigma$]=0.075), Kormendy
($L \propto\re^{1.58}$, rms[$\log\re$]=0.1), and edge-on
FP\footnote{Note that the slopes of the three considered scaling laws
are mutually consistent within the errors, i.e., the combination of
the FJ and the Kormendy relations gives the adopted edge-on FP best
fit. Similar results are obtained using the K-band relations of Pahre
et al. (1998).}  ($\re\propto\sigma^{1.51} I_e^{-0.77}$,
rms[$\log\re]= 0.049$) relations in the z-band as given by Bernardi et
al. (2003a,b); we also consider the $\mbh -\sigma$ relation (Ferrarese
\& Merritt 2000; Gebhardt et al. 2000). An important issue of the
present analysis is then how to map, for each galaxy model, the two
sets $(\mstar,\rv,\sigv)$ and $(L,\re,\sigma)$.

Because we are not using $N$-body simulations -- where under the
assumption of a constant mass-to-light ratio $\ML$ the relation
between virial and "observed" properties is known (e.g., see NLC03) --
we adopt a conservative approach, and we assume a mass-dependent
structural \emph{weak homology} of our galaxies compatible with the FP
tilt (e.g., see BCD02): in practice, we "force" the models to stay on
the edge-on FP, and then we check if and how the FJ and Kormendy
relations are preserved.  This assumption is well founded, both
observationally and theoretically.  In fact, it is known that the
edge-on FP is characterized by a tilt, i.e., by a systematic trend of
the ratio between the stellar mass-to-light ratio $\ML$ and the virial
parameter $\Kv$
\begin{equation}
{\ML\over\Kv}\propto L^{-\frac{\alpha+2\beta}{\alpha}}
\re^{\frac{2+\alpha+4\beta}{\alpha}},
\end{equation}
where $G\mstar=\Kv\re\sigma^2$, and identity above holds for the
general expression of the edge-on FP $\log\re=\alpha\log\sigma
+\beta\log I_e +\gamma$ (curiously, we note that while in the B-band
all the tilt depends on luminosity (BCD02), in the present case and in
the K-band it is almost due only to $\re$, with $\ML/\Kv\propto
L^{0.02}\re^{0.28}$ (e.g., see Treu 2001).  Unfortunately, a definite
answer about the origin and the physical driving parameter(s) of the
FP tilt is still missing. It is however known that structural weak
homology could be able to produce the whole (or a large part of) the
FP tilt. In particular, Sersic (1968) models provide a remarkably good
description of the light-profiles of elliptical galaxies (e.g., see
Caon et al. 1993, Graham \& Colless 1997), with the Sersic index $n$
increasing with galaxy luminosity and spanning the range of values
required by equation (20) to reproduce the FP tilt (e.g., see Ciotti,
Lanzoni \& Renzini 1996, Ciotti \& Lanzoni 1997, BCD02). Note that an
increase of $n$ with galaxy mass was also found in $N$-body
simulations of major mergers (NLC03). Thus, in this paper we introduce
weak homology by assuming that for a galaxy characterized by the pair
$(\rv,\sigv)$, the observables $\re$ and $\sigma$ are given by
\begin{equation}
{\rv\over\re}\simeq {250.26+7.15 n\over 77.73+n^2},
\end{equation}
and 
\begin{equation}
{\sigma\over\sigv}\simeq {24.31+1.91 n +n^2\over 44.23+0.025 n +0.99 n^2}.
\end{equation}
The two relations above, where $\sigma$ is the luminosity wheighted
projected velocity dispersion within $\re/8$, hold with very good
accuracy for one-component, isotropic Sersic models with $2\lsim n
\lsim 12$ (Ciotti 1991, Ciotti \& Lanzoni 1997, Ciotti \& Bertin
1999). From equations (21) and (22) the corresponding virial
coefficient $\Kv(n)=(\rv/\re)\times (\sigv/\sigma)^2$ is easily found
(see also CBD02).  How a specific value of $n$ is assigned to a given
galaxy model is described in the following Sections.

\subsection{Equal mass merging}
\label{sec:eqM}

\begin{figure}[ht]
\plotone{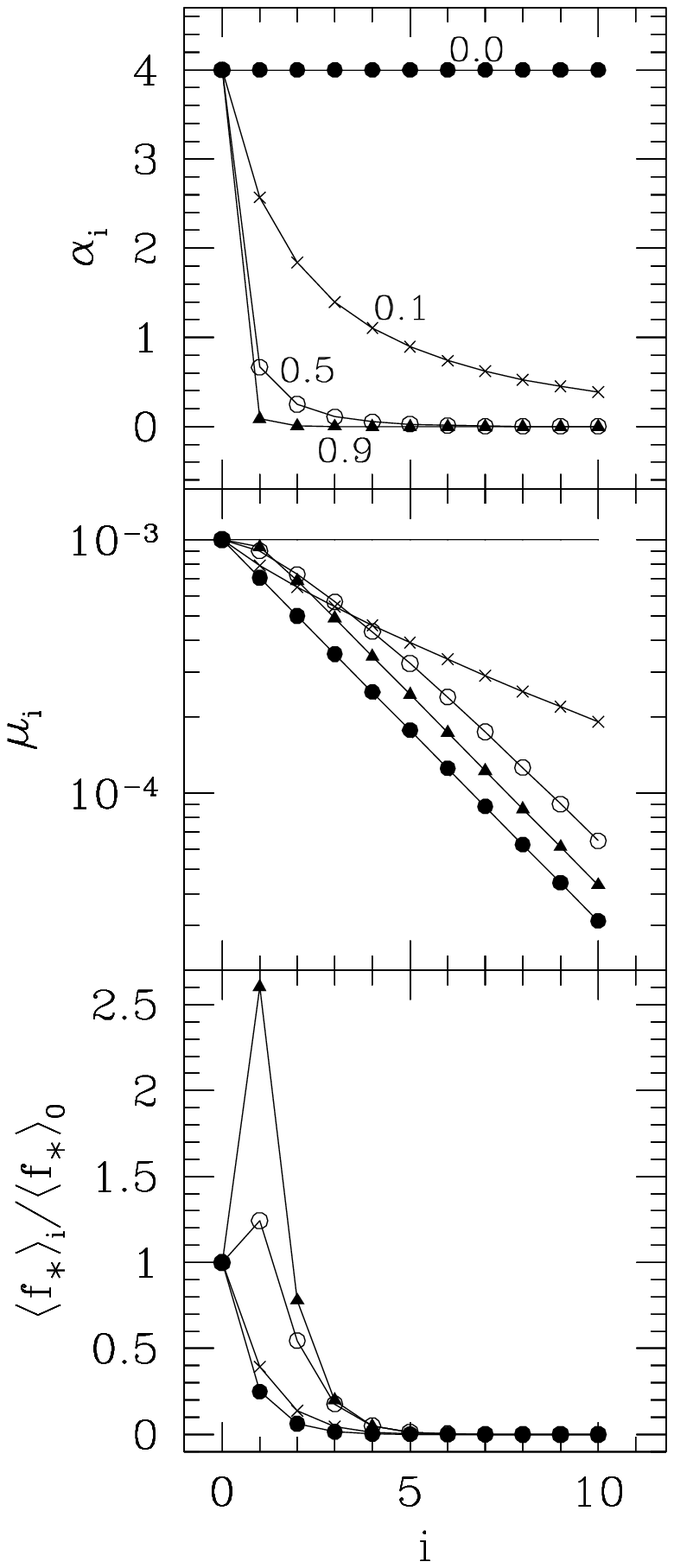}
\caption{The evolution of the gas-to-star mass ratio $\alpha_i$ (top
panel), of the BH-to-star mass ratio $\mu_i$ (middle panel), and of
the stellar mean phase-space density $\fstarmean_i
\equiv\langle \rhostar \rangle_i/\sigma_{vi}^3$ (bottom panel) in the
case of 10 successive equal-mass parabolic mergers, and for different
values of the dissipation parameter $\eta$ (0, solid dots; 0.1,
crosses; 0.5 empty circles; 0.9, solid triangles). In all the merging
sequences, $\alpha_0\equiv M_{g0}/M_{\ast 0}=4$, and
$f=\mu_0=10^{-3}$.  Maximum radiative efficiency ($p=2$) is assumed
for the BH coalescence law in the middle panel ($p=1$ case is
represented by the horizontal line).}
\label{fig:alpimuifei}
\end{figure}

In order to illustrate the effect of repeated dry and wet mergers on a
population of elliptical galaxies, in this Section we start our
analysis presenting the idealized case of a merging hierarchy of equal
mass spheroids, extending the analysis of CvA01 to the dissipative
case.  The seed galaxies (zeroth-order generation) are identical
systems characterized by a stellar mass $\mstarz$, a gas mass
$\mgasz=\alpha_0\mstarz$, a central BH mass $M_{BH0} =\mu_0\mstarz$, a
virial radius $r_0$, and a virial velocity dispersion $\sigma_0$. A
galaxy of the $i$-th generation is the merger-product of two galaxies
of $(i-1)$-th generation, so the equations
(\ref{eq:mstar})-(\ref{eq:sigv}) can be written in recursive form.
The solution of the gas mass equation (\ref{eq:mgas}), that in the
present case reads $M_{gi+1}=2(1-\eta -f\eta)M_{gi}$, is
\begin{equation}
\mgasi= (2\,q)^i \mgasz, \quad\quad q\equiv 1-\eta-f\eta,
\label{eq:mgasi}
\end{equation}
so that for $q\le1/2$ the gas mass is a steadily decreasing quantity along
the merging hierarchy. The stellar mass equation (\ref{eq:mstar}) becomes 
 $M_{\ast i+1}=2\,M_{\ast i} +2\eta M_{gi}$, and from equation
(\ref{eq:mgasi}) we
obtain
\begin{equation}
\mstari= 2^i \left(1+ \alpha_0{1-q^i\over 1+f}\right)\mstarz,
\label{eq:mstari}
\end{equation}
while the gas-to-star mass fraction at stage $i$ is given by
\begin{equation}
\alpha_i=\frac{\alpha_0 \, q^i\,(1+f)} {1+f+\alpha_0 \,(1-q^i)};
\label{eq:alpi}
\end{equation}
at variance with $M_{gi}$, $\alpha_i$ is a decreasing function of $i$
independently of the value of $q$.  In Fig.\ref{fig:alpimuifei} we
show the evolution of $\alpha_i$ along sequences of ten equal-mass
mergers, starting from gas-dominated seed galaxies ($\alpha_0=4$) and
for different values of $\eta$ (with $f=10^{-3}$): according to
equation (\ref{eq:mstari}) the stellar mass increases by a factor
$\sim 10^3$ for $\eta=0$ and of $\sim 5\times 10^3$ for
$\eta=0.9$. The horizontal solid line $\alpha_i=\alpha_0$ represents
the dry merging case ($\eta=0$; note that we call ``wet'' a merging in
which gas dissipation is active: a gas rich merger with $\eta=0$ is in
practice a dry merger).  When significant dissipation is present,
$\alpha_i$ dramatically decreases in the first mergers for the
combined effects of gas depletion and the associated stellar mass
increase. Only for values of $\eta$ as low as $\sim 0.1$ a more gentle
evolution is produced.

The BH mass evolution equation (\ref{eq:mbh}), $M_{BH i+1} = 2^{1/p}
M_{BH i} +2f\eta M_{g i}$, is solved with the aid of equation
(\ref{eq:mgasi}) and reads
\begin{equation}
M_{BHi} =2^{i/p} M_{BH0}\times\cases{\displaystyle{
          1+{f\eta\alpha_0\over\mu_0}
            {2^{i(1-1/p)}q^i -1\over q-2^{1/p-1}},\quad\quad q\ne 2^{1/p -1}};
                               \cr\displaystyle{ 
          1+{f\eta\alpha_0\over\mu_0}\;2^{1-1/p}\,i, \quad\quad q= 2^{1/p -1}};
          }
\label{eq:mbhi}
\end{equation}
the explicit formula of the BH-to-star mass ratio $\mu_i$ (equation
[\ref{eq:mu}]) can be derived from equation (\ref{eq:mstari}) and the
equation above.  The evolution of $\mu_i$ is shown in the middle panel
of Fig.\ref{fig:alpimuifei} for the maximally radiative case $p=2$ and
for fixed $f=\mu_0=10^{-3}$: while the Magorrian relation is preserved
by construction in case of classical BH merging ($p=1$, solid
horizontal line), in the extreme ($p=2$) case $\mu_i$ decreases for
increasing galaxy mass, even though fresh gas is added to the BH at
each merging in proportion to the stellar mass increase.  Thus, in
order to preserve the Magorrian relation when $p>1$, an increasing
fraction $f\eta$ of the available gas must be accreted on the BH as
the merging hierarchy proceeds, increasing the AGN activity.  When the
progenitor spheroids are gas rich, high values of $\eta$ may initially
compensate the decrease of $\mu$ due to gravitational radiation;
however, after a few mergers these galaxies run out of gas, and the
final values of $\mu$ are even lower than in the less dissipative
$\eta=0.1$ case.

From equations (\ref{eq:sigp}) and (\ref{eq:rp}) we finally obtain the
relations between the virial velocity dispersion and the virial radius
of the progenitors and of the new galaxy:
\begin{equation}
{\sigma_{v\,i}^2\over \sigma_{v\,i-1}^2} = 1+
{\eta(1+2f)\alpha_{i-1}\over 1+(1-f\eta)\alpha_{i-1}}, 
\label{eq:sigvi} 
\end{equation} 
\begin{equation}
{r_{v\,i}\over r_{v\,i-1}} = {2[1+(1-f\,\eta)\alpha_{i-1}]^2\over
(1+\alpha_{i-1})\left[1+\alpha_{i-1}+\eta(1+f)\alpha_{i-1}\right]}.
\label{eq:rvi} 
\end{equation} 

As expected, $\sigv$ is larger (and $\rv$ is smaller) in the wet than
in the dry merging case: for example, $\sigma_{v\,i}^2 \sim
\sigma_{v\,i-1}^2\, (1+\eta)$ and $r_{vi} \sim 2\,
r_{v\,i-1}/(1+\eta)$ in the limit of $\alpha_{i-1} \gg 1$.  Figure
\ref{fig:alpimuifei} shows how gas dissipation may produce a
non-monothonic behavior of the quantity $\fstarmean_i \equiv\langle
\rhostar\rangle_i/\sigma_{vi}^3 = 3\,\mstari/(8\,\pi r_{vi}^3
\sigma_{vi}^3)$, which is often considered an estimate of the
phase-space density.  In particular, while $\fstarmean_i$ decreases as
$\fstarmean_0/4^i$ in equal mass dry merging, in highly dissipative
gas rich mergers the increase of $\langle \rhostar\rangle$ dominates
over the increase of $\sigv^3$, and $\fstarmean_i\sim\fstarmean_{i-1}
(1+\eta\alpha_{i-1}) \,(1+\eta)^{3/2}/4$.  From the previous formula
one would then conclude that an increase of the phase-space density is
limited to exceptionally gas rich mergers, but this is not correct. In
fact $\langle f_\ast\rangle$ is based on virial quantities, that by
their nature refer to global scales: an increase of the phase-space
density in the galactic central regions can be produced by the
\emph{localized} dissipation of a smaller amount of gas.

\begin{figure}[ht]
\plotone{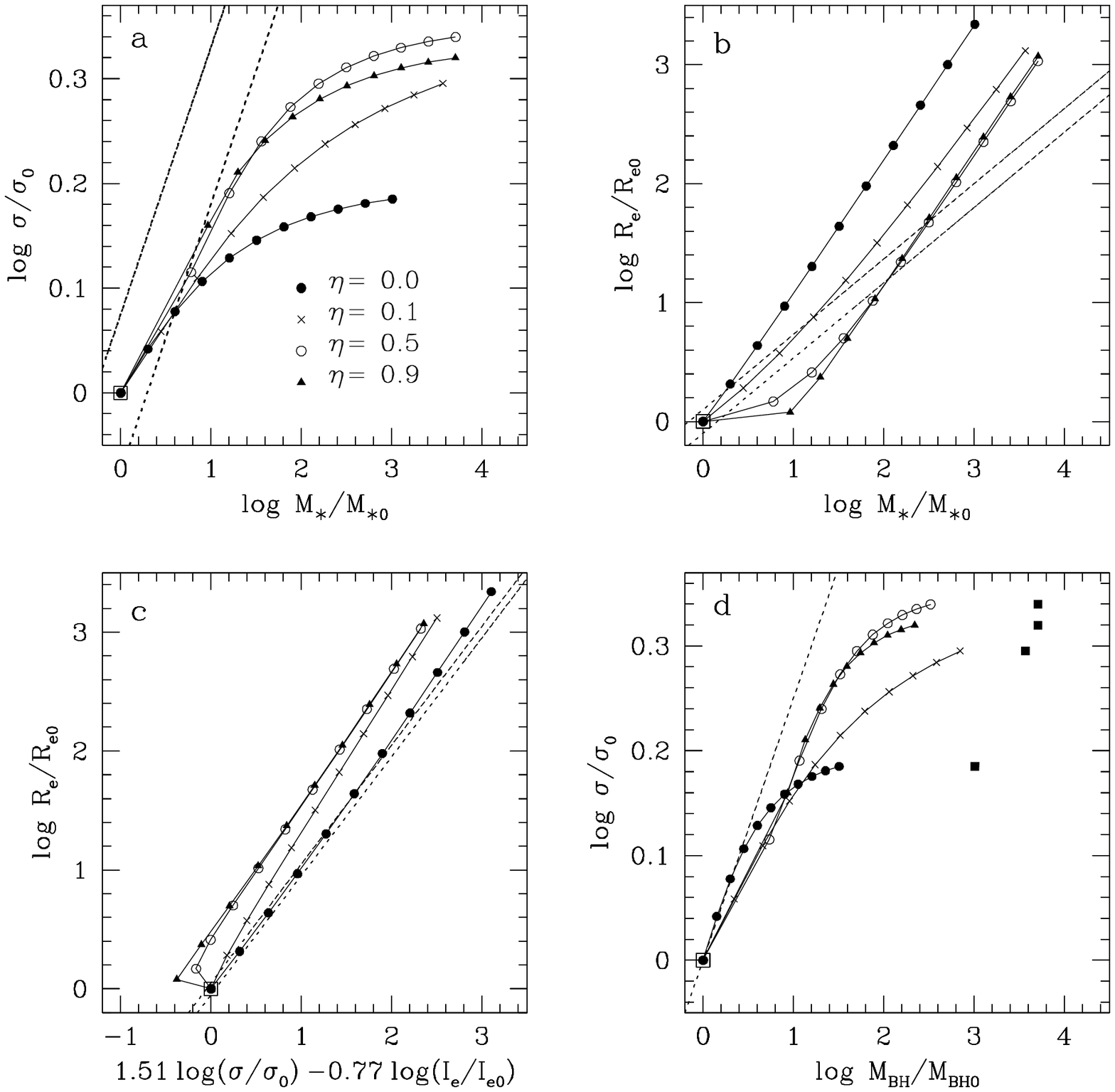}
\caption{The models of Fig.\ref{fig:alpimuifei} in the scaling
relation planes.  Dotted lines in panels $a$--$c$ represent the
observed scaling relations in the z-band with their 1-rms scatter. In
panel $d$ the $\mbh -\sigma$ relation is plotted without scatter, and
$p=2$ is assumed for the BH coalescence formula, while black squares
mark the position of the last merger product if $p=1$. See Section
\ref{sec:eqM} for details.}
\label{fig:eqM_screl}
\end{figure}

In Fig.\ref{fig:eqM_screl} we plot the representative points of the
same models of Fig.\ref{fig:alpimuifei} in the $(\mstar,\sigma)$,
$(\mstar,\re)$, $(\mstar,\re ,\sigma)$, and $(\mbh,\sigma)$
planes. These plots, under the assumption of the same stellar
mass-to-light ratio $\ML$ for all models, correspond to the FJ (panel
a), Kormendy (panel b), and FP (panel c) planes.  The assumption of a
constant value for $\ML$ is made less severe by the comparing the
models to the observed scaling laws in the z-band (dashed lines),
where metallicity effects on $\ML$ are reduced with respect to bluer
wavelenghts. Merging induced structural weak homology is imposed by
assuming that the seed galaxies are Sersic $n=2$ models in accordance
with the observed light profiles of low-luminosity ellipticals, and
that $n$ increases by 1 in each merging, as shown by numerical
simulations (NLC03). In this way, the final range of values spanned by
$n$ is between 2 and 12, consistently with observations.  In practice,
for assigned values of $r_0$ and $\sigma_0$ of the seed galaxies, from
equations (21) and (22) with $n=2$ we obtain their $\re$ and
$\sigma$. We also assume that the seed galaxies are placed at the
lower end of the various scaling laws represented in Fig. 2. The equal
mass merging formula (27)-(28) are then applied, and the new virial
radius and velocity dispersion are mapped in the corresponding $\re$
and $\sigma$ again from equations (21) and (22) with $n=3$, and so on.

From Fig.2c it is apparent how the FP tilt is well reproduced by the
models corresponding to dry mergers (solid dots). The adopted
prescription for weak homology is relevant here: in fact, it is easy
to prove that if the models were plotted by using $\rv$ and $\sigv$
instead of the fiducial $R_e$ and $\sigma$, they would be placed along
a line of slope $-1/\beta\sim 1.3$ (for a surface-brightness
coefficient $\beta=-0.77$ in the FP expression) instead of 1.  Figure
2c also shows that highly dissipative wet mergers are initially
displaced from the FP, but they again move along lines almost parallel
to the edge-on FP as soon as a large fraction of gas is converted into
stars.  These simple considerations indicate that the final position
of a galaxy in the FP space as a consequence of merging is sensitive
to the physical processes involved, as already discussed by Bender,
Burstein \& Faber (1993).  At variance with the edge-on FP, neither
the FJ nor the Kormendy relations are reproduced: in particular, while
velocity dispersions are too low, effective radii are too large.
Again, note that this inconsistency would be exacerbated when plotting
$\rv$ and $\sigv$: for example, solid dots in Fig.\ref{fig:eqM_screl}a
would be aligned on a horizontal line, while in
Fig.\ref{fig:eqM_screl}b they would be placed on the line
$\rv\propto\mstar$. From Fig.\ref{fig:eqM_screl}d it is finally
apparent how the $\mbh -\sigma$ relation is also failed, expecially in
the classical merging case.  Remarkably, for $p=2$ the mass loss due
to emission of gravitational waves maintains the BH mass nearer to the
observed relations the the classical merging case. In general, wet
mergers are in better agreement with the FJ and Kormendy relations
than dry mergers (in a way dependent on the specific value of $\eta$),
due to the shrinking of $\rv$ and the increase of
$\sigv$. Unfortunately, in the present framework we cannot evaluate
the galaxy non-homology induced by gas dissipation, that can be
investigated only with N-body$+$gas numerical simulations such those
of Robertson et al. (2006a), and so weak homology is just imposed with
the same prescription as for dry mergers. In any case, the preliminary
analysis of this Section is consistent with the idea that the FJ and
Kormendy relations are stronger tests for merging than the edge-on FP,
as already clearly shown by numerical simulations (e.g., see
NLC03; Boylan-Kolchin et al. 2005, 2006).

\section{The simulations}
\label{sec:simu}

In this Section we extend the previous investigation to the study of
the effects of repeated parabolic merging on a \emph{population} of
elliptical galaxies. The merging spheroids are extracted by means of
Monte-Carlo simulations from different samples of seed galaxies, and
the properties of the resulting galaxies are determined by using the
relations derived in Section \ref{sec:models}.  The motivation for these
simulations is the fact that equal-mass merging maximizes (minimizes)
the effects on the radius (velocity dispersion) of the resulting
objects, while mergers of galaxies spanning a substantial range of
masses, sizes and velocity disperions not only are more realistic, but
also could lead to less dramatic effects on the scaling relations.

In particular, we focus on two schemes designed to explore and
quantify the impact of dry and wet merging on the formation and
evolution of elliptical galaxies.  In the first scheme the seed
galaxies span only a narrow mass range (a factor of $\sim 5$): in this
case we then study whether massive ellipticals and the observed
scaling relations can be \emph{produced} by repeated mergers of
low-mass spheroidal systems.  In the second scheme the seed
ellipticals follow the observed scaling relations over their whole
observed mass range ($\sim 10^3$), and so we explore whether repeated
merging events preserve or destroy these relations.  For sake of
completeness, and also to check the robustness of the results obtained
with the Monte-Carlo simulations, we finally conduct a third set of
experiments in which the merging histories are described by Press \&
Schechter (1974) merger trees.

How the mass, virial radius and velocity dispersion of the seed
galaxies, as well as their effective radius and central velocity
dispersion, are assigned in each experiment is described in the
corresponding Sections. In all cases, however, the initial mass of the
SMBH obeys the Magorrian relation with $\mu_0=10^{-3}$.

\subsection{Merging small seed galaxies}
\label{sec:narrow}

In this first scheme, once two spheroids are extracted from the
initial population (made of 1000 objects), they are merged together,
and the properties of the merger end-product are computed as described
in Section \ref{sec:models}. The two progenitors are then removed from
the seed galaxy population, while the new object is added to it; the
procedure is repeated until the largest produced galaxies are $\sim
10^3$ times more massive than the smallest seed galaxy in the original
sample.  This may require up to 10-12 mergers, $\sim $7-10 of which
being major mergers (i.e., merging in which the stellar mass ratio of
the progenitors is in the range $0.3-3$; e.g., see Kauffmann,
Guiderdoni \& White 1994).  The initial population of seed galaxies is
obtained by random extraction (with the von Neumann rejection
technique) of the stellar mass $\mstar$ from the SDSS z-band galaxy
luminosity function (Blanton et al. 2001), under the assumption of
constant stellar mass-to-light ratio $\ML$. Finally, the mass ratio of
the most massive to the less massive galaxy in the sample is taken to
be 5. In case of wet mergers, the {\it total} (stars+gas) mass is the
quantity which is extracted. For each galaxy mass, the corresponding
central velocity dispersion $\sigma$ is fixed according to the z-band
FJ, and the effective radius $\re$ is assigned from the FP in the
z-band (Bernardi et al. 2003a,b). Due to the restricted mass range,
all the galaxies are assumed to be $n=2$ Sersic models, and so their
virial radius $\rv$ and virial velocity dispersion $\sigv$ can be
easily calculated. We then apply the rule that in major mergers the
Sersic index of the resulting galaxy is $n=1+\max(n_1,n_2)$, where
$n_1$ and $n_2$ are the Sersic indices of the progenitors. In minor
mergers, the Sersic index instead keeps the same value of the more
massive galaxy. Note that this is a quite conservative assumption,
because in NLC03 it was found that in head-on minor mergers $n$
actually {\it decreases}, producing galaxies that fall outside the
edge-on FP.

\begin{figure}[ht]
\plotone{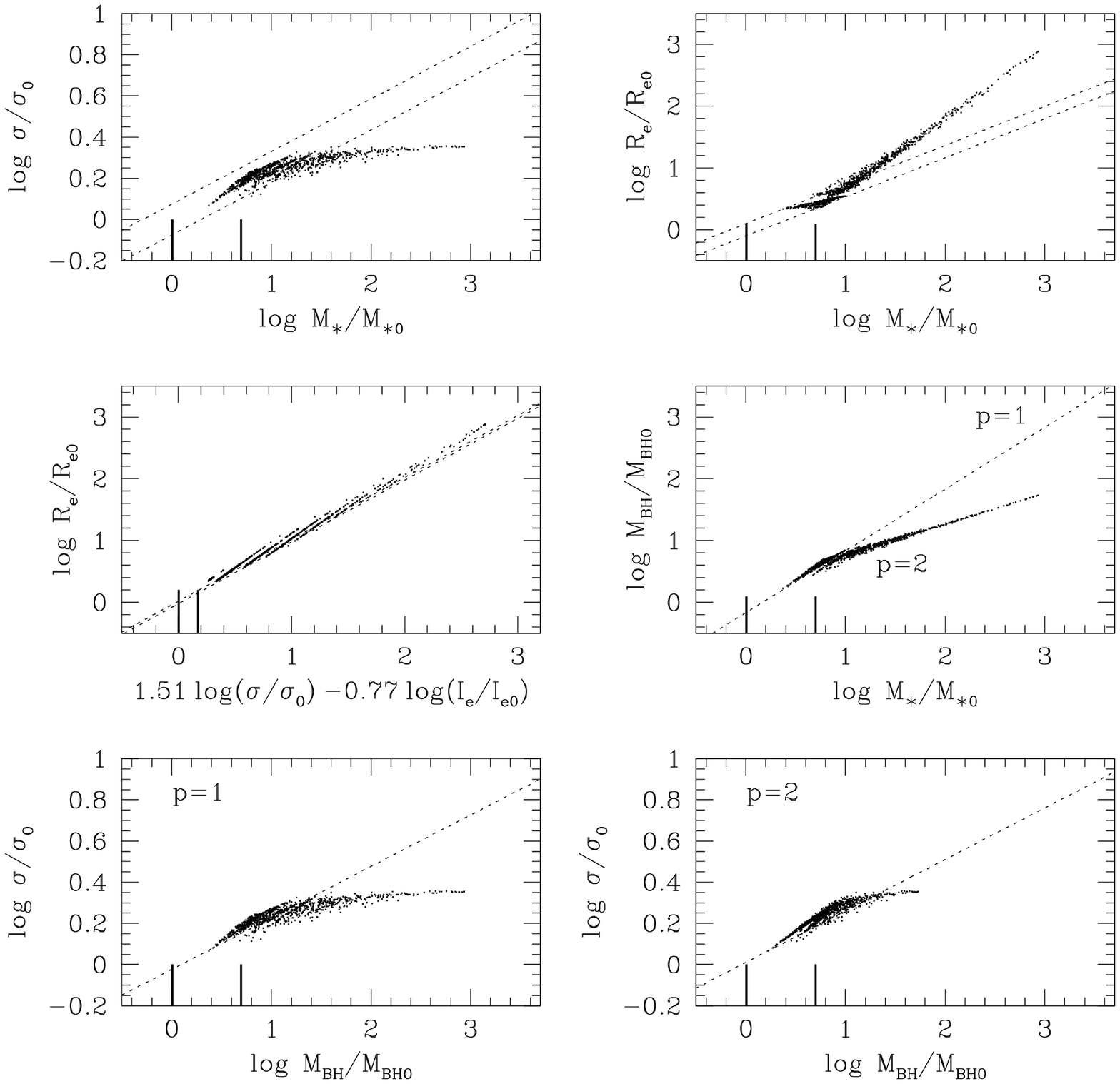}
\caption{Synthetic scaling relations produced by parabolic dry mergers. Seed
galaxies span a limited mass range (indicated by the heavy vertical ticks)
and random re-merging events are repeated until a factor $10^3$ increase in
mass is reached (see text for details).  Dotted lines represent the observed
scaling relations, as in Fig.\ref{fig:eqM_screl}.  All quantities are
normalized to the properties of the lowest mass seed galaxy.
\label{fig:dry_smallM}}
\end{figure}

\begin{figure}[ht]
\plotone{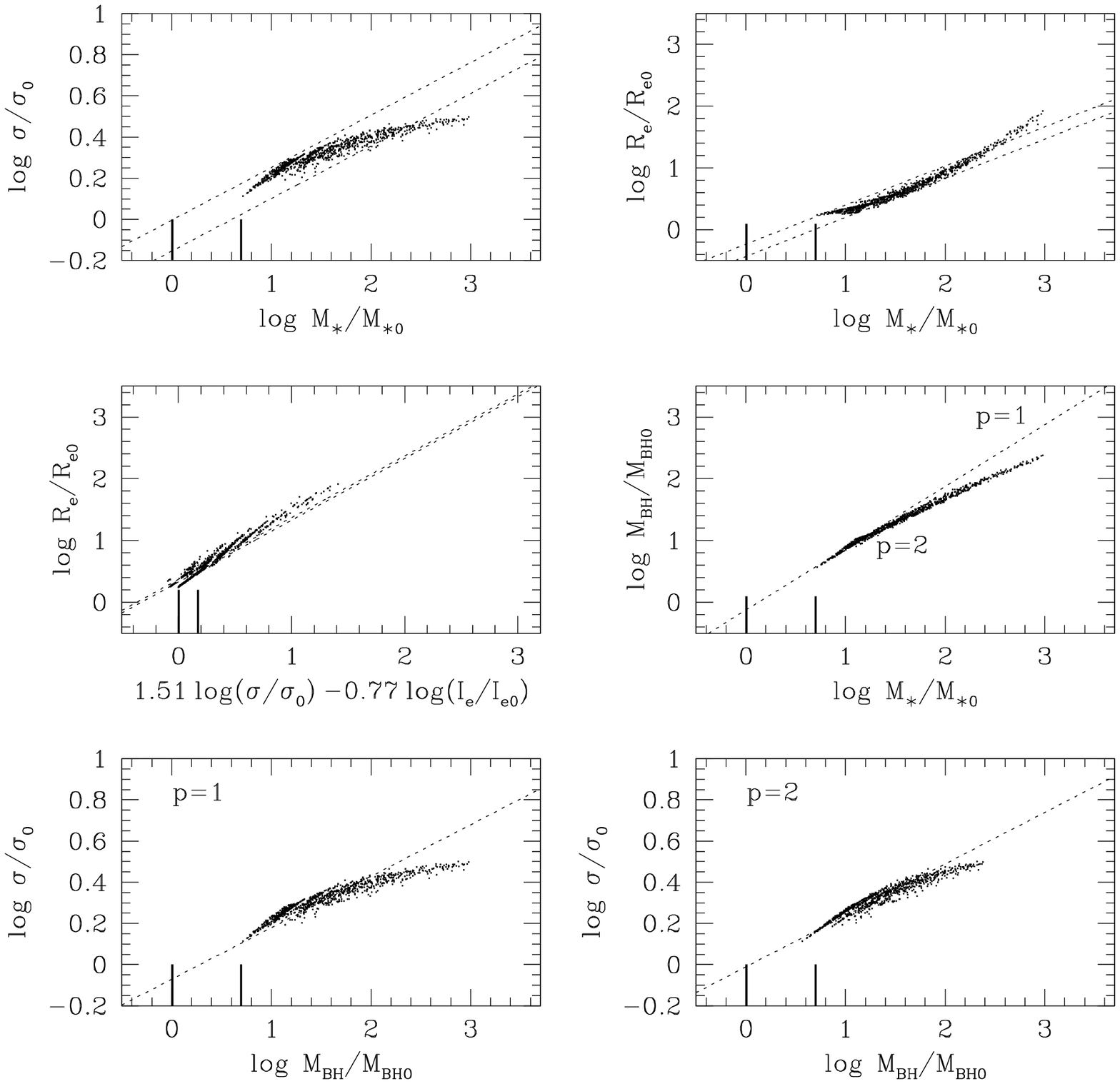}
\caption{As in Fig.\ref{fig:dry_smallM}, but for the wet merging of
initially gas-rich galaxies: $\alpha_0=4$ and $\eta=0.3$.
\label{fig:wet_smallM}}
\end{figure}

Figures \ref{fig:dry_smallM} and \ref{fig:wet_smallM} show the results
in the cases of dry ($\alpha_0=0$) and dissipative gas rich
($\alpha_0=4, \eta=0.3$) parabolic merging, respectively; the mass
interval spanned by the progenitors is indicated by the two vertical
ticks, the end-product positions are represented by the dots, and the
observed scaling laws are represented by the dotted lines.  Figure
\ref{fig:dry_smallM} reveals that massive ellipticals cannot be formed
by parabolic dry mergers of low-mass spheroids only, because they
would be characterized by exceedingly large $\re$ and almost mass
independent $\sigma$, in agreement with the results of CvA01 and
NLC03, and with the conclusions of Section \ref{sec:models}.  In fact,
when galaxies reach a mass $\sim 10$ times larger than that of the
largest seed galaxies, all the seed galaxy population can be
considered made by equal mass objects, and the considerations of
Section 2.2 apply.

Figure \ref{fig:wet_smallM} shows the results in the case of wet
merging of gas dominated ($\alpha_0=4$) galaxies. As expected, mergers with 
gas dissipation produce more realistic objects than dry mergings and,
remarkably, the observed scaling laws are satisfied (even though with
a large scatter) by the new galaxies, up to a mass increase of a
factor of $10^2$ with respect to the smallest seed galaxies. However,
new galaxies characterized by a mass increase factor $\gsim 10^2$ are
mainly formed by mergers of gas poor galaxies that already experienced
several mergers, and so they deviate from the observed scaling laws as
the galaxies in Fig.\ref{fig:dry_smallM}.

More quantitatively, the models plotted in Fig.\ref{fig:dry_smallM}
deviate from the observed $\mbh-\sigma$ and FJ relations by more than
$1-\sigma$ when their (logarithmic) mass increase is $\gsim 1.4$ and
$\gsim 2.4$, respectively, the larger mass value allowed by the FJ
being due to its larger scatter.  The mean gas-to-star mass ratio for
the deviating models is $\alpha\lsim 0.5$ (even though several models
with a lower $\alpha$ are still consistent with the two relations
considered).  Quite obviously, these values depend on the initial
amount of gas: for example, when starting with $\alpha_0=10$ the
models are incompatible with the observed $\mbh-\sigma$ for a
logarithmic increase of the stellar mass of $\gsim 2.2$, and for
$\alpha\lsim 0.3$. We note, however, that the populated region in the
edge-on FP is reduced for increasing $\alpha_0$, as can be seen by
comparing the model distributions in Figs.~3 and 4.

This first exploration therefore reveals that parabolic merging of low
mass galaxies only is unable to produce elliptical galaxies obeying
the observed scaling laws, even when allowing for structural weak
homology in a way consistent with the edge-on FP. However gas
dissipation plays an important role in gas rich merging and remarkably
the resulting elliptical galaxies appear to be distributed as the
observed scaling relations, as far as enough gas is available.  Quite
obviously, the problem of the compatibility of the properties of such
merger-products with other key observations, such as the
color-magnitude and the metallicity-velocity dispersion relations, and
the increasing age of the spheroids with their mass (e.g., see Renzini
2006; Gallazzi et al. 2006) cannot be addressed in the framework of
this paper.

\subsection{Merging "regular" galaxies}

In the second scheme, the masses of the seed galaxies span the full
range covered by ordinary ellipticals ($\sim 10^3$), and their
characteristic size and velocity dispersion follow the observed
scaling relations.  The mass, effective radius, and central velocity
disperion of each seed galaxy are assigned as described in Section
3.1; however, due to the large mass range spanned in the present case,
the models cannot be caracterized by the same value of the Sersic
index, if they are placed on the edge-on FP. For this reason a Sersic
index is assigned to each seed galaxy by solving for $n$ the equation
$\Kv (n)= G\mstar/\re\sigma^2$; in turn, from the knowledge of $n$ we
obtain the values of $\rv$ and $\sigv$ needed in the merging
scheme. For simplicity we restrict our study to major mergers only,
increasing by 1 the larger Sersic index characterizing each merging
pair. Finally, for consistency with the imposed scaling laws (which
hold for present-day gas poor spheroids), we focus on dry merging
only. In this Section we then study the effect of merging on already
established scaling laws.

\begin{figure}[ht]
\plotone{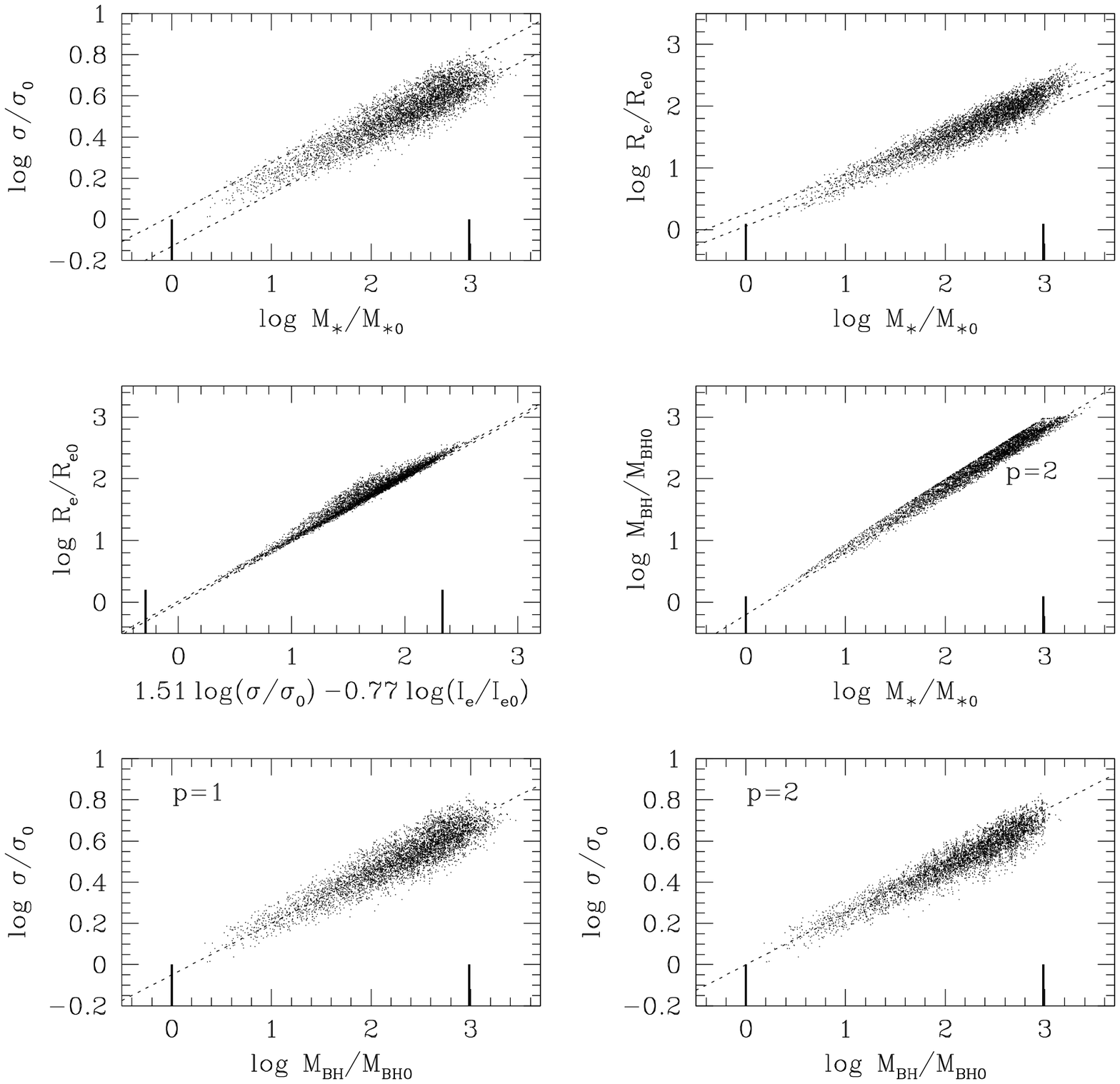}
\caption{Synthetic scaling relations for the merger-products of up to 6 dry
major mergers of galaxies extracted from a population that follows the
observed scaling laws. Lines are as in Fig.\ref{fig:eqM_screl} and all
quantities are normalized to the properties of the lowest-mass seed galaxy.
\label{fig:dry_allM}}
\end{figure}

In practice, once a galaxy is chosen, a second galaxy with a mass
ratio to the first in the range $0.3-3$ is extracted from the seed
population, and then the two galaxies are merged. As in the other
cases, the intrinsic galaxy properties are transformed into their
"observational" counterparts by using equations (21) and (22). The
procedure is repeated by selecting a third galaxy from the initial
population, and so on for a total of 6 major mergers.  The positions
in the observational planes of 1000 galaxies (at all stages of the
merging hyerarchy) are shown in Fig.\ref{fig:dry_allM}. The main
result is that now, at variance with the narrow mass range
experiments, the scaling laws remain almost unaffected by the merging,
both in their slope and scatter. In particular, note how the
$\mbh$-$\sigma$ relation (with $p=2$) is preserved, even though we are
in dry merging regime. The only detectable deviations from the
observed scaling laws, for the same reasons already discussed in
Section \ref{sec:narrow}, are found for ellipticals with masses larger
than the most massive galaxies in the original sample (marked by the
two vertical ticks in Fig.\ref{fig:dry_allM}).

Why mergers do preserve so well the scaling relations?  The reason is
simple: by construction in a population of galaxies spanning the whole
mass range observed today and distributed according to the observed
scaling laws, mergers in general involve a "regular" elliptical, with
realistic values of $\re$ and $\sigma$. These mergings act as a
``thermostat'', maintaining values of $\re$ in the observed range and
increasing the virial velocity dispersion, thus contributing to
preserve the scaling laws. Only when the produced galaxies are so
massive that no regular galaxies of comparable mass are available, the
new merger products deviate more and more from the scaling laws. This
behavior becomes extreme in the case of repeated mergers in a galaxy
population spanning a restricted mass range, as discussed in Section
\ref{sec:narrow}. Thus, our analysis confirm that while the elliptical
galaxy scaling laws (and so elliptical galaxies) cannot be produced by
the merging of low mass spheroids only (as already pointed out by
e.g., CvA01, NLC03, Evstigneeva et al. 2004), these relations once
established by some other mechanism, are robust against merging.

\subsection{Cosmological merger trees}
\label{sec:tree}

We conclude our study by presenting a set of numerical experiments
aimed at investigating the evolution of galaxies with a merging
history obtained from the extended Press \& Schechter (1974) formalism
in a standard $\Lambda$CDM cosmology. The details of the realization
of the merger trees are given in Volonteri, Haardt \& Madau (2003),
while the ensemble from which they have been extracted is in
accordance with the Jenkins (2001) modified P\&S formula. In
particular, we selected a set of 20 merger trees tracing the merger
history, from $z=5$ to $z=0$, of a halo with mass $\simeq
10^{13}M_\odot$ at present time.  Since the mass resolution in the
merger tree scales as $M_{\rm res}=10^{10}\,(1+z)^{-3.5} M_\odot$, it
follows that $M_{\rm res}$ is always $\lsim 5$\% of the main halo mass
in the merger hierarchy. This wide range of masses allows for both
minor and major mergers in the tree at all redshifts.  
that the {\it statistical} discrepancies between the DM 
function derived from the Press \& Schechter formalism and 
numerical simulations (e.g., Jenkins et al. 2001) are not 
the present context, where we focus on the growth of a 

In practice, we applied at each merging event in a given tree the
relations derived in Section \ref{sec:models}, both for the dry and
the wet ($\alpha_0=4, \eta=0.3$) cases, for a total of 40 simulations.
The virial radius $\rv$ of each seed halo (that we arbitrarily
identify with a galaxy) is now defined as the radius of the sphere
characterized by mean mass density $\Delta_{\rm vir}\,\rho_{\rm crit}$
(where $\rho_{\rm crit}$ is the critical density for closure at
redshift $z$, and $\Delta_{\rm vir}$ is the density contrast at
virialization\footnote{For the assumed cosmology this can be
approximated by $\Delta_{\rm vir}\simeq 178\Omega^{0.45}$ (Eke,
Navarro, \& Frenk 1998).}).  This definition of virial radius is not -
strictly speaking - identical to the standard dynamical relation
(\ref{eq:Wstar}).  However, in Lanzoni et al. (2004) it was shown that
the two definitions of $\rv$ are in nice agreement, and so we also
define the halo (galaxy) virial velocity dispersion from the virial
theorem $G M=\rv\sigv^2$.  The properties of the merger end-product
are determined according to the dry or wet relations, while those of
the secondary galaxy involved in each subsequent event follow the
cosmological virial relations. The weak homology trend is added to the
models by assigning a Sersic index $n=2$ to the main halo at $z=5$
(which is assumed to be placed on the reference scaling laws) , and
increasing it by 1 in each major merger; in minor mergers $n$ is
maintained constant.

For simplicity, in Fig.6 we show the FJ, Kormendy and FP planes for
just two out of the 40 simulations, being the behavior of galaxy
models in all the merger trees almost identical. Note that at variance
with Figs.3,4, and 5, here the points constitute an evolutionary
sequence, representing the successive positions of the main halo
during its mass accretion history. From the comparison with
Figs.\ref{fig:dry_smallM} and \ref{fig:wet_smallM} it is apparent that
deviations from the slope of the FJ and Kormendy relations are less
strong than in the previous cases, while the evolutionary tracks of
the growing halos move parallel to the edge-on FP plane.

The fact that also in the merger tree exploration the slopes of the
FJ, Kormendy, and FP relations are preserved is not surprising, as it
is easily explained when combining the results of previous Sections
with the fact that now the "galaxies" involved in the mergings are
provided by the cosmological setting, in which
$M\propto\rv^3\propto\sigv^3$. Thus, the determinant factor of success
is again the availability of galaxies with a virial radius and
velocity dispersion increasing with the halo mass, a property that
cannot be produced by parabolic mergings of small systems only, but it
is the natural consequence of the substantially different phenomenon
of negative energy collapses (see Introduction).

Note that the accordance with observations would be in fact even
better than the results shown in Fig. 6. According to the hierarchical
merging picture, the number of mergers an elliptical galaxy
experiences in its lifetime (efficient mergers, i.e., those with time
scales shorter than the Hubble time) is much smaller than the number
of halo mergers in a cosmological merger tree, as only a small
fraction ($\lsim 30\%$) of them leads to galaxy mergers, once the
finite time needed for merging is taken into account (see
Fig.\ref{fig:effmerg}). In fact, dynamical friction appears to be very
efficient (i.e., with a decay timescale shorter than the Hubble time)
only for mergers with a mass ratio of the progenitors $\gsim 0.1$
(Taffoni et al. 2003), while satellites in intermediate mass ratio
($0.01 - 0.1$) suffer severe mass losses by the tidal perturbations
induced by the gravitational field of the primary halo, and this
progressive mass loss further increases the decay time.  The lightest
satellites are almost unaffected by orbital decay, so they survive and
keep orbiting on rather circular, peripheral orbits.

\begin{figure}[ht]
\plotone{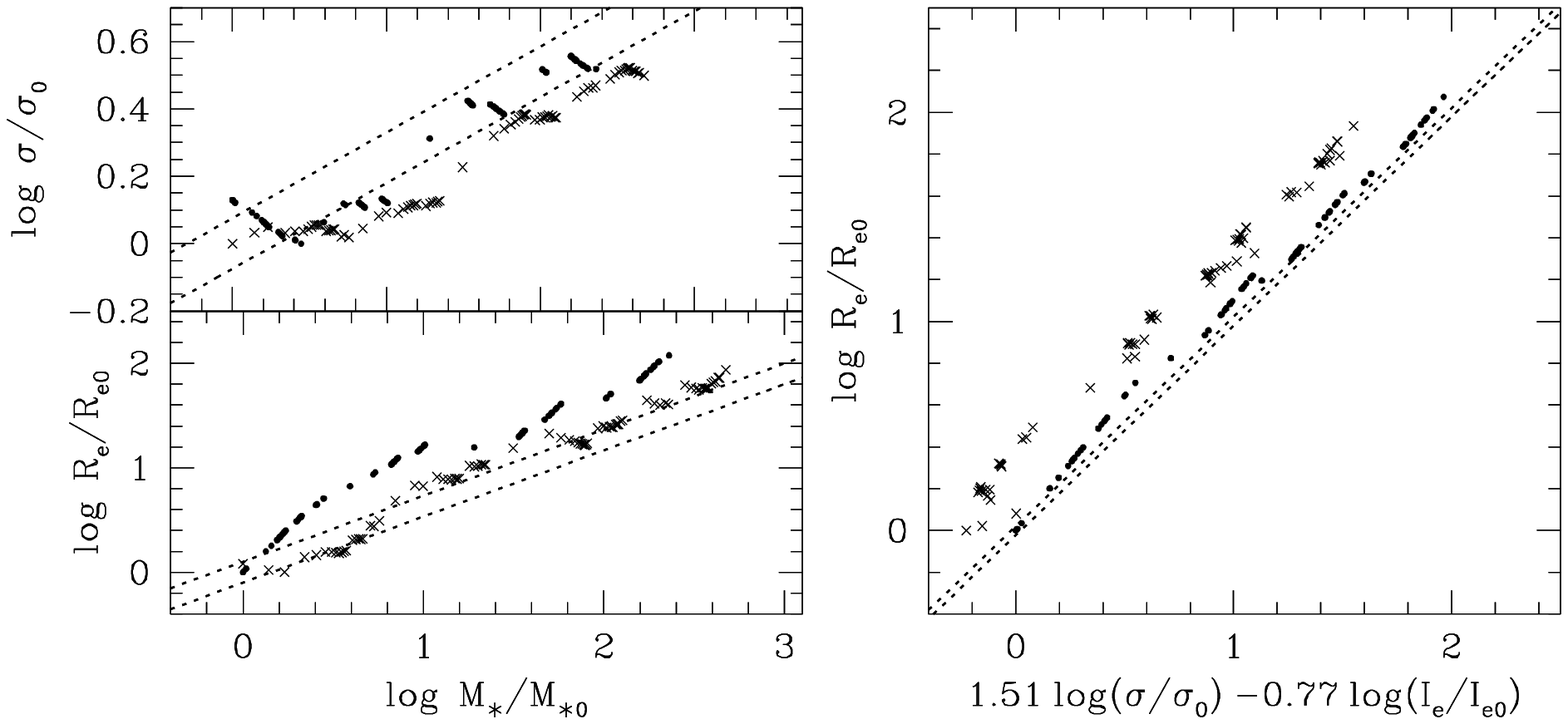}
\caption{Evolutionary sequences of a main halo in the scaling law planes,
according to a Press \& Schechter merger tree, in the dry (dots) and wet
($\alpha_0=4$ and $\eta=0.3$; crosses) cases.
\label{fig:tree}}
\end{figure}
%
\begin{figure}\centering
\plotone{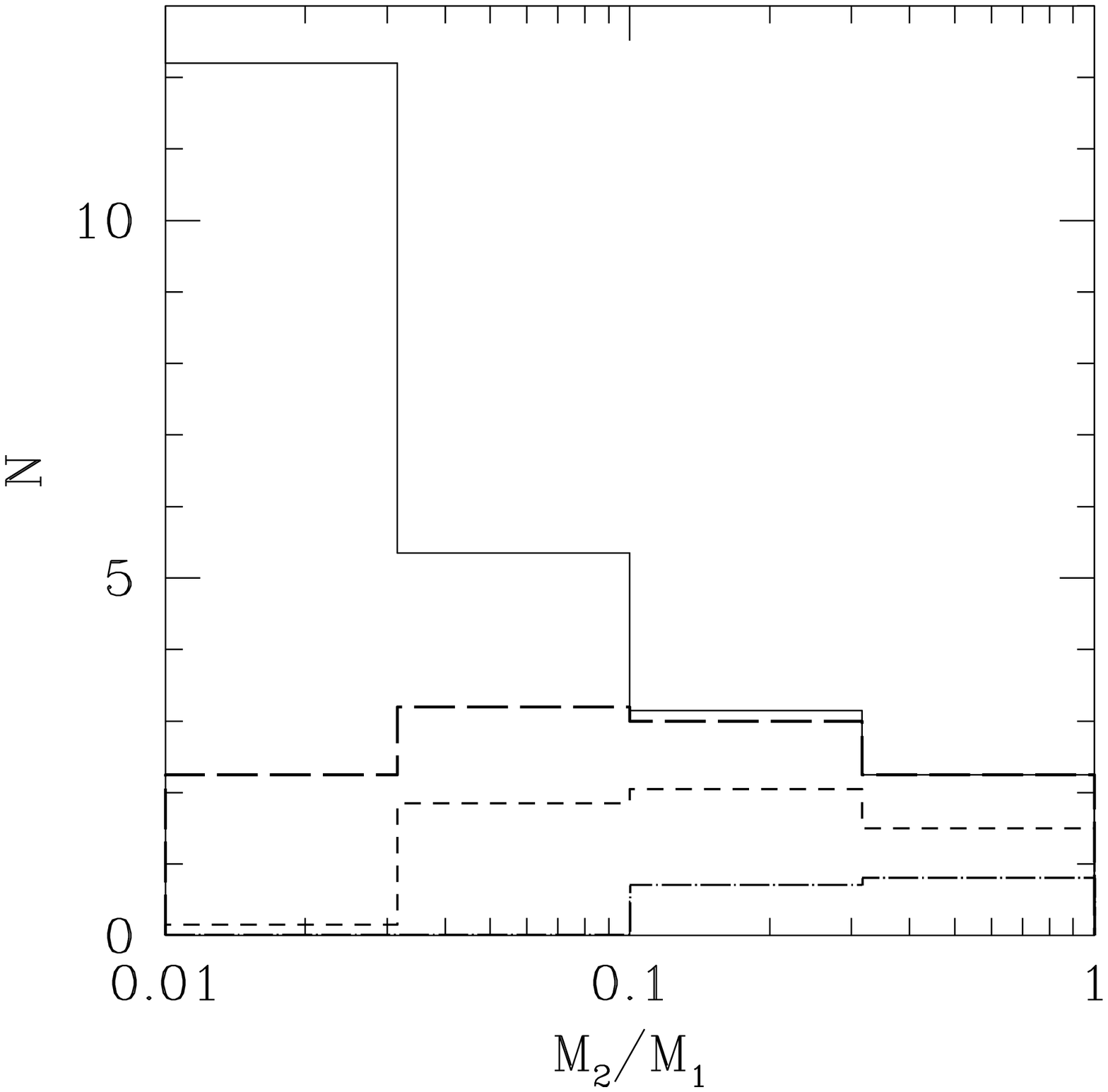}
\caption{Number of mergers per logarithmic secondary-to-primary mass ratio,
extracted from the merger history of a $M_0=10^{13}M_\odot$ halo at $z=0$,
and averaged over 20 merger trees. The total number of mergers experienced by
the halo at $z<3$ is shown by the solid histogram.  $z<3$. The number of
\emph{efficient} mergers (see the text for the definition) experienced by the
same halo at different redshift is shown by the long-dashed (for $z<3$),
short-dashed (for $z<2$), and dot-dashed (for $z<1$) histograms.
\label{fig:effmerg}}
\end{figure}

\section{Discussion and conclusions} 
\label{sec:concl} 

With the aid of a scheme based on very simple physical arguments we
investigated the influence of dry and wet merging on the formation and
evolution of elliptical galaxies, focusing on the origin and
robustness of some of their scaling laws.  In particular, by using
analytical arguments and numerical simulations we showed that massive
elliptical galaxies cannot be formed by (parabolic) merging of low
mass spheroidal galaxies, even in presence of substantial gas
dissipation, and allowing for the helpful effects of structural weak
homology. However the observed scaling laws of elliptical galaxies,
once established by galaxy formation, are robust against merging. More
specifically, our findings can be summarized as follows:

1) Parabolic dry merging in a population of low mass spheroids leads
to massive ellipticals that cannot be simultaneously placed on the
Kormendy, FJ and edge-on FP relations. For example, forcing the
end-products to stay on the edge-on FP, the FJ and Kormendy relations
are failed by massive galaxies, with deviations increasing with galaxy
mass. This behavior was predicted in CvA01 and confirmed by high
resolution numerical simulations (NLC03, Boylan-Kolchin et al. 2005,
2006).  For example Boylan-Kolchin et al. (2006), in a series of
dissipationless merging simulations of galaxies in cosmologically
motivated orbits, found that the merging end-products, while
preserving the edge-on view of the FP, can present significant
differences in the FJ and Kormendy relations. This because the
variations in the resulting $M_*-\re$ are compensated by corresponding
variations in the $M_*-\sigma$ relation, so that the {\it projections}
of the FP - but not the edge-on FP itself - should provide a powerful
way to investigate the assembly history of massive elliptical
galaxies.

2) Parabolic wet merging in the same population of low mass
progenitors lead to galaxies in much better agreement with the
observed scaling relations, as long as enough gas for dissipation is
available.  In particular, the resulting $\mbh$-$\sigma$ relation is
in better agreement with the observed one, also in the case of
significant mass loss (via gravitational waves) of the coalescing BHs.
Significant deviations from the observed scaling laws are however
expected for massive galaxies. Similar conclusions were reached by
sophisticated $N$-body plus hydrodynamical simulations of merging of
disk galaxies. For example, Kazantzidis et al. (2005) found that
merging disk galaxies constructed to obey the $\mbh-\sigma$ relation
move relative to it depending on whether they undergo a dissipational
or dissipationless merger. In particular, remnants of dry mergings
tend to move away from the mean relation, showing the role of gas-poor
mergers as a possible source of scatter. In addition, Robertson et al
(2006b) studied the development of the $\mbh-\sigma$ over cosmic time
with a large set of hydrodynamical simulations of galaxy mergers that
include star formation and feedback from the growth of the central BH,
and found that the $\mbh-\sigma$ relation is created through coupled
BH and spheroid growth (via star formation) in galaxy mergers.

3) Parabolic dry mergers in a population of galaxies following the
observed scaling laws over the full mass range populated today by
stellar spheroids (or following the scaling laws of dark matter halos
predicted by the current cosmological scenario), preserve the
Kormendy, FJ, and edge-on FP remarkably well. The reason of this
behavior is rooted in the availability in the merger population of
galaxies with velocity dispersion increasing with galaxy mass.
Remarkably, Robertson et al. (2006a) found evidences that dry merging
of spheroidal galaxies at low redshift is expected to maintain the FP
relation {\it imprinted by gas-rich merging during the epoch of rapid
spheroid and central BH growth at high redshift}, when the progenitors
were characterized by gas fractions $\gsim 30\%$ and efficient gas
cooling was allowed in the simulations.

Thus, points 1) and 2) above suggest that ellipticals cannot be
originated by parabolic merging of low mass spheroids only, even in
presence of substantial gas dissipation. In addition, point 3), when
considered together the cosmologically imprinted scaling laws of dark
matter halos, and the several appealing features of dissipationless
collapse end-products (see Introduction), support the idea that
elliptical galaxies formed in a process similar to the monolithic
collapse, even though their structural and dynamical properties are
compatible with a limited number of dry mergers (we note that the same
conclusion has been reached also from the study of color profiles
in early-type galaxies, as described in Wu et al. 2005).

The possibility that monolithic collapse and successive merging are
just the leading physical processes at different times in galaxy
evolution, and that they are both important for galaxy formation, is
perhaps indicated also by a "contradictory" and often overlooked
peculiarity of massive ellipticals In fact, while the Kormendy
relation dictates that the mean stellar density of galaxies decreases
for increasing galaxy mass (a natural result of parabolic dry
merging), the normalized light profiles of elliptical galaxies becomes
steeper and their metallicity increases at increasing galaxy mass (as
expected in case of significant gas dissipation).  Thus, the
present-day light profiles of ellipticals could represent the fossil
evidence of the impact of both the processes; quite obviously, this
problem cannot be addressed in the framework adopted in this paper.
It would be very interesting to extend the Robertson et al. (2006a) and
Naab et al. (2006) analysis to the study of dissipative collapses in
cosmologically motivated dark matter halos, thus extending the
investigation of Nipoti et al. (2006) towards the very early phases of
galaxy formation.

\acknowledgements {We thank the anonymous Referee for useful comments
that improved the presentation of the paper. L.C. acknowledges the
warm hospitality of the Theoretical Astrophysics Division of
Harvard-Smithsonian Center for Astrophysics, where a large part of
this work was carried out.  Lars Hernquist, Jerry Ostriker, Brant
Robertson and Tjeerd van Albada provided insightful comments. L.C. is
supported by MIUR, grant CoFin2004.}

\end{document}